\documentclass{aa}  

\usepackage{graphicx}
\usepackage{txfonts}
\usepackage[colorlinks=true,citecolor=blue]{hyperref}
\usepackage{threeparttable} %for table note
\usepackage[normalem]{ulem}

\begin{document}

   \title{Exploring the atmosphere of GJ 1132 b with CRIRES+}

%   \subtitle{}

   \author{E. Palle
          \inst{1,2}
          \and
          F. Yan
          \inst{3}
          \and
          G. Morello
          \inst{4}
          \and
          M. Stangret
          \inst{5}
          \and
          M.R. Swain
          \inst{6}    
          \and
          J. Orell-Miquel
          \inst{1,2} 
          \and
          P. Miles-Paez
          \inst{7}
          \and
          R. Estrela
          \inst{6}
          \and
          T. Masseron
          \inst{1,2}
          \and
          G. Roudier
          \inst{6}
          \and
          P.B. Rimmer
          \inst{6}
    }

% G. Morello \inst{4} \orcid{0000-0002-4262-5661}
% J. Orell-Miquel \inst{1,2} \orcid{0000-0003-2066-8959}
% P. A. Miles-P\'aez \inst{5} \orcid{0000-0003-2446-8882}
% Raissa Estrela \inst{5}  0000-0002-4006-6755.
% Paul B. Rimmer   0000-0002-7180-081X
% Thomas Masseron 0000-0002-6939-0831
% GAel Roudier 0000-0002-7402-7797
% Monika Stangret 0000-0002-1812-8024
% MArk Swain 0009-0001-4487-7299

   \institute{Instituto de Astrof\'isica de Canarias (IAC), 38200 La Laguna, Tenerife, Spain \\
   \email{epalle@iac.es}
   \and
   Deptartamento de Astrof\'isica, Universidad de La Laguna (ULL), 38206 La Laguna, Tenerife, Spain
   \and
   Department of Astronomy, University of Science and Technology of China, Hefei 230026, China\\
   \email{yanfei@ustc.edu.cn}
   \and
   Instituto de Astrof\'isica de Andaluc\'ia (IAA-CSIC), Glorieta de la Astronom\'ia s/n, 18008 Granada, Spain
   \and
   INAF – Osservatorio Astronomico di Padova, Vicolo dell’Osservatorio    \and
   California Institute of Technology, NASA Jet Propulsion Laboratory
   \and
   Centro de Astrobiolog\'ia, CSIC-INTA, Camino Bajo del Castillo s/n, 28692 Villanueva de la Ca\~nada, Madrid, Spain
%   \and
%    \textcolor{red}{Missing full address}
%   \and
%    \textcolor{red}{Missing full address}
%   \and
%    \textcolor{red}{Missing full address}
   }

   \date{Received September 15, 1996; accepted March 16, 1997}

% \abstract{}{}{}{}{} 
% 5 {} token are mandatory
 
  \abstract
   {With a mass, radius, and mean density similar to Earth's, the rocky planet GJ 1132 b is the first truly small planet for which an atmosphere detection was proposed. If confirmed, ultra-reduced magma outgassing is the only mechanism capable of producing HCN and H$_2$O in large enough quantities to match the HST observations. The proposed atmosphere detection, however was challenged by reanalysis of the same HST data by different teams. Recent JWST observations returned ambiguous results due to the unaccounted for variability seen between two different visits.  Here we report the analysis of three CRIRES+ transit observations of GJ 1132 b in order to determine the presence or absence of He I, HCN, CH$_4$, and H$_2$O in its atmosphere. We are unable to detect the presence of any of these species in the atmosphere of GJ 1132 b assuming a clear, H$_2$-dominated atmosphere, although we can place upper limits for the volume mixing ratios of CH$_4$, HCN, and H$_2$O using injections tests and atmospheric retrievals. These retrieved upper limits show the capability of CRIRES+ to detecting chemical species in rocky exoplanets, if the atmosphere is H$_2$ dominated. The detection of the atmospheres of small planets with high mean molecular weight, and the capability to distinguish between the variability introduced by stellar activity and/or the planetary atmosphere will require high-resolution spectrographs in the upcoming extremely large telescopes.}


   \maketitle
%
%-------------------------------------------------------------------

\section{Introduction}

The atmospheric characterization of temperate Earth-size rocky planets is a necessary step to accomplish the ultimate goal of searching for possible biosignatures in the atmospheres of habitable planets. Planets around M dwarfs are particularly suited for this endeavour because of the favorable planet/star contrast ratio and the fact that planets in the habitable zone of those stars have relatively short orbital periods. 

While we are still far from reaching this goal, the atmospheres of several warm ($T_\mathrm{eq} < 1000K$) mini-Neptunes and super-Earths have been explored using the Hubble Space Telescope (HST). Excluding the TRAPPIST-1 system planets, where more HST orbits may be required, there have been 12 published transmission spectra for planets with $R < 3 R_{\oplus}$; of these five were reported as flat spectra (GJ 1214 b – \citet{Kreidberg2014}; HD 97658 b – \citet{Knutson2014}; L98-59 b, c and d - \citet{Damiano2022,Zhou2022,Zhou2023}) and 7 have been reported with spectral features (55 Cnc e – \citet{Tsiaras2016}; K2-18 b –\citet{Benneke2019,Tsiaras2019}; HD 3167 c – \citet{Guilluy2021,Mikal-Evans2021}; LHS 1140 b – \citet{Edwards2021}; GJ 1132 b – \citet{Swain2021}; TOI-270 d - \citet{Mikal-Evans2023}; GJ 9827 d - \citet{Roy2023}).
%ORIGINAL-START 5 have been reported with spectral features (55 Cnc e – \citet{Tsiaras2016}; K2-18 b –\citet{Benneke2019,Tsiaras2019}; HD 3167 c – \citet{Guilluy2021,Mikal-Evans2021}; LHS 1140 b – \citet{Edwards2021}; GJ 1132 b – \citet{Swain2021}).%ORIGINAL-END
More recently the JWST has become a much more powerful tool to explore the atmospheres of small planets.
Still, the detection of an atmosphere in transmission often requires several transits to reach a sufficient signal-to-noise ratio (S/N). Even then, there is always the possibility of clouds or a very tenuous atmosphere returning a very precise but flat spectrum or the possibility of stellar contamination effects. The first examples are the recent JWST results on LHS 457b by \citet{Lustig2023} and GJ 486b by \citet{Moran2023}. Such results are inconclusive on the existence or nature of a planetary atmosphere. However, robust atmospheric detections were obtained for K2-18 b \citep{Madhusudhan2023} and TOI-270 d \citep{Holmberg2024,Benneke2024arxiv}, also solving previous degeneracies between H$_2$O and CH$_4$. This body of work  shows that the characterization of small planet atmospheres has begun. However, because small planets have the capacity to lose, regenerate, and totally transform their atmospheres, their study will profoundly deepen and broaden the exoplanet field.

With a mass ($\sim 1.6 M_{\oplus}$), radius ($\sim 1.1 R_{\oplus}$) and mean density ($\sim 6.3 g/cm^3$) similar to Earth's, the rocky planet GJ 1132 b \citep{Berta-Thompson2015, Bonfils2018} is the  smallest planet for which an atmosphere might have been detected. 
While GJ 1132 b is significantly hotter ($T_{eq} \approx 530K$) than Earth, it is one of a handful of terrestrial transiting planets with accurate mass determination, and therefore of great interest for comparative planetology. GJ 1132 b is believed to have lost any possible primordial H$/$He envelope \citep{Schaefer2016}, although H from the primordial envelope could have been dissolved in the mantle \citep{Chachan2018, Kite2019}. GJ 1132 b has also been identified as a candidate for reestablishment of an atmosphere by mantle outgassing \citep{Kite2020} and the possibility of a steam atmosphere and the presence of a substantial O$_2$ layer have been theoretically explored \citep{Schaefer2016}. 

%GJ 1132b orbits a K= 8.3 magnitude M3.5 star, and has a transit duration (T1-T4) of 47 min. 

Early ground-based multiband photometry suggested the presence of H$_2$O and/or CH$_4$ to explain the absorption features observed in the z and K bands \citep{Southworth2017}. However, another ground-based optical spectrum did not show any evident atmospheric modulation \citep{Diamond-Lowe2018}. HST/STIS searches for the Ly$\mathrm{\alpha}$ transit of GJ 1132 b were not successful neither, leading the authors to conclude that any original H/He envelope had been lost \citep{Waalkes2019}.

Using data from HST WFC3, the transmission spectrum of GJ 1132 b was obtained. \citet{Swain2021} analyzed the data and reported the spectral signatures of aerosol scattering, HCN, and CH$_4$ in a low mean molecular weight atmosphere, an unexpected composition that was the subject of an extensive modeling effort. A combination of atmospheric loss, thermal/photochemical, and geochemical modeling, identified ultra-reduced magma outgassing as a plausible mechanism capable of producing HCN and CH$_4$ in quantities matching the observations \citep{Swain2021}. 
%This interpretation, if correct has broad implications for future detection of terrestrial planet atmospheres, and confirmation of the results is urgently needed. 
However, a reanalysis of the same WFC3 data obtained discrepant results. Both \citet{Mugnai2021} and \citet{Libby-Roberts2022} reported no detections of molecular signatures and their transmission spectrum was best fit with a flat-line model. Their results suggest that the planet does not have a clear primordial, hydrogen-dominated atmosphere, and that instead, GJ 1132 b could have either a cloudy hydrogen-dominated atmosphere, a very enriched secondary atmosphere, or an airless/tenuous atmosphere.

More recently the JWST observed two transits of GJ 1132 b with the NIRSpec G395H configuration. However, rather than solving the issue, the JWST data exhibit substantial differences between the two visits \citep{May2023}. 
The first transit was consistent with a H$_2$O-dominated atmosphere with CH$_4$ and N$_2$O, or with stellar contamination from unocculted starspots, while the second transit returned a featureless transmission spectrum. The authors could not identify the source of this differences/variability. \citet{Xue2024} observed GJ 1132 b's thermal emission (secondary eclipse) with the Mid-Infrared Instrument Low-Resolution Spectrometer (MIRI/LRS) on the JWST, and concluded that the planet likely does not have a significant atmosphere. Thus the existence and composition of GJ 1132 b's atmosphere remains unsolved. Here we present our attempt to explore this elusive planet's atmosphere using ground-based high-resolution spectroscopy.

\section{Observations and data reduction}
We observed three transits of GJ 1132 b on December 14, 2021, February 9, 2022 and January 22, 2023  with the CRIRES+ high-resolution spectrograph mounted on the Very Large Telescope \citep{Dorn2023}. The first transit was carried out in Y-band with the Y1029 wavelength setting to search for the Helium 1083\,nm lines. The other two transits were performed in H-band with the H1559 configuration to search for molecules such as HCN, $\mathrm{CH_4}$, and H$_2$O. We used the 0.2\,\arcsec slit to achieve a high spectral resolution (R $\sim$ 120\,000). The observations were performed in staring mode, and we used the adaptive optics system. The exposure time was set to 300\,s. Each night's observation lasted for approximately 3 hours. For the first observation we obtained 33 spectra with 10 in-transit spectra. For the the other two nights, we obtained 32 spectra, which consists of 9 spectra during transit and 23 spectra out of transit for each night (see Table~\ref{obs_log}).

%We observed two transits of GJ 1132b on February 9, 2022 and January 22, 2023  with the CRIRES$^+$ high-resolution spectrograph mounted on the Very Large Telescopes \citep{Dorn2023}. The observation was performed with the H1559 configuration to search for HCN, $\mathrm{CH_4}$ and H$_2$O. We used the 0.2\,\arcsec slit to achieve a high spectral resolution (R $\sim$ 120\,000). The observation was performed in staring mode, and we used the adaptive optics system. The exposure time was set to 300\,s. Each night's observation lasted for approximately 3 hours, and we obtained 32 spectra, which consists of 9 spectra during transit and 23 spectra out of transit for each night. 

We reduced the raw spectra using the CRIRES+ pipeline. The pipeline handles calibrations including the dark subtraction, flat-fielding and wavelength calibrations. The science frame was reduced with the \texttt{cr2res\underline{~}obs\underline{~}staring} command. 
The reduced spectrum of the Y1029 wavelength setting has 9 spectral orders, covering the wavelengths from 950\,nm to 1120\,nm.
For the H1559 wavelength setting, the spectrum consists of 8 spectral orders from 1460\,nm to 1820\,nm. We discarded the reddest order of the the H1559 spectrum because it has low flux with strong telluric absorption. The wavelength solution provided by the pipeline has a slight offset around a few km\,s$^{-1}$. We therefore refined the wavelength solution with telluric lines using the \texttt{molecfit} software \citep{Smette2015}. We further calculated the spectral drifts between consecutive exposures during the observing night. The drifts are below 0.1 pixel during the observation and we subsequently corrected the drift for each spectrum. 

For each exposure frame, we normalized the spectra order-by-order and then merged the spectra into a one-dimensional spectrum. We discarded the data points with low S/N. About 10\% of the data points were discarded. These discarded points are mostly at wavelengths of strong telluric lines.

%
%                                             Simple A&A Table
%-----------------------------------------------------------
\begin{table}
\small
\caption{Observation logs.}             % title of Table
\label{obs_log}      % is used to refer this table in the text
\centering                          % used for centering table
\begin{threeparttable}
        \begin{tabular}{l c c c c c}        % centered columns (4 columns)
        \hline\hline    \noalign{\smallskip}               % inserts double horizontal lines
        ~ &        Date &   Exposure time & $N_\mathrm{spectra}$  &  Wavelength \\     % table heading
        \hline  \noalign{\smallskip}                  % inserts single horizontal line
  Night 1 & 2021-12-14 &    300 s & 33 & Y1029  \\ 
  \hline  \noalign{\smallskip}                                %inserts single line
  Night 2 & 2022-02-09  &    300 s & 32 & H1559  \\ 
\hline  \noalign{\smallskip}  
  Night 3 & 2023-01-22  &   300 s & 32 & H1559 \\ 
\hline
        \end{tabular}
%\tablefoot{
%\tablefoottext{a}{These values are from the header of the FITS file, which was measured at 550\,nm. The seeing in the infrared is expected to be better than these values.} }
\end{threeparttable}      
\end{table}
%-----------------------------------------------------------

\section{Results}
\subsection{\ion{He}{i} transmission spectroscopy}

%                                                Two column figure
%----------------------------------------------------------- S_vib
   \begin{figure}
   \centering
   \includegraphics[width=0.48\textwidth]{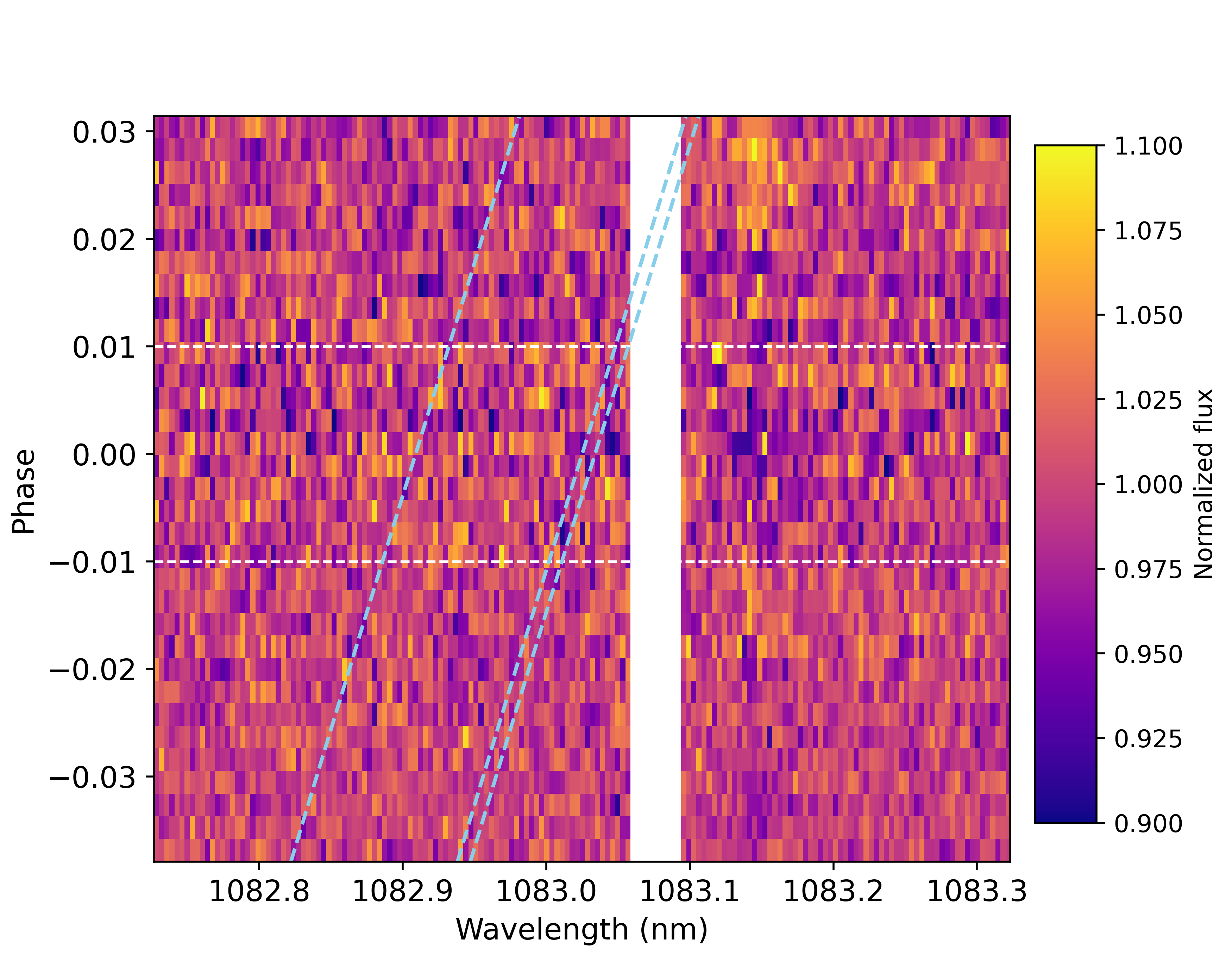}
      \caption{Transmission spectral matrix around the \ion{He}{i} 1083\,nm lines. The three blue dashed lines denote the expected position of the planetary \ion{He}{i} absoption lines. The horizontal white dashed lines indicate the beginning and end of the transit.  The white region is the masked pixels around the strong telluric OH line.
       The spectra are represented in the stellar rest frame.}
         \label{He-matrix}
   \end{figure}
%-----------------------------------------------------------

%                                                Two column figure
%----------------------------------------------------------- S_vib
   \begin{figure}
   \centering
   \includegraphics[width=0.48\textwidth]{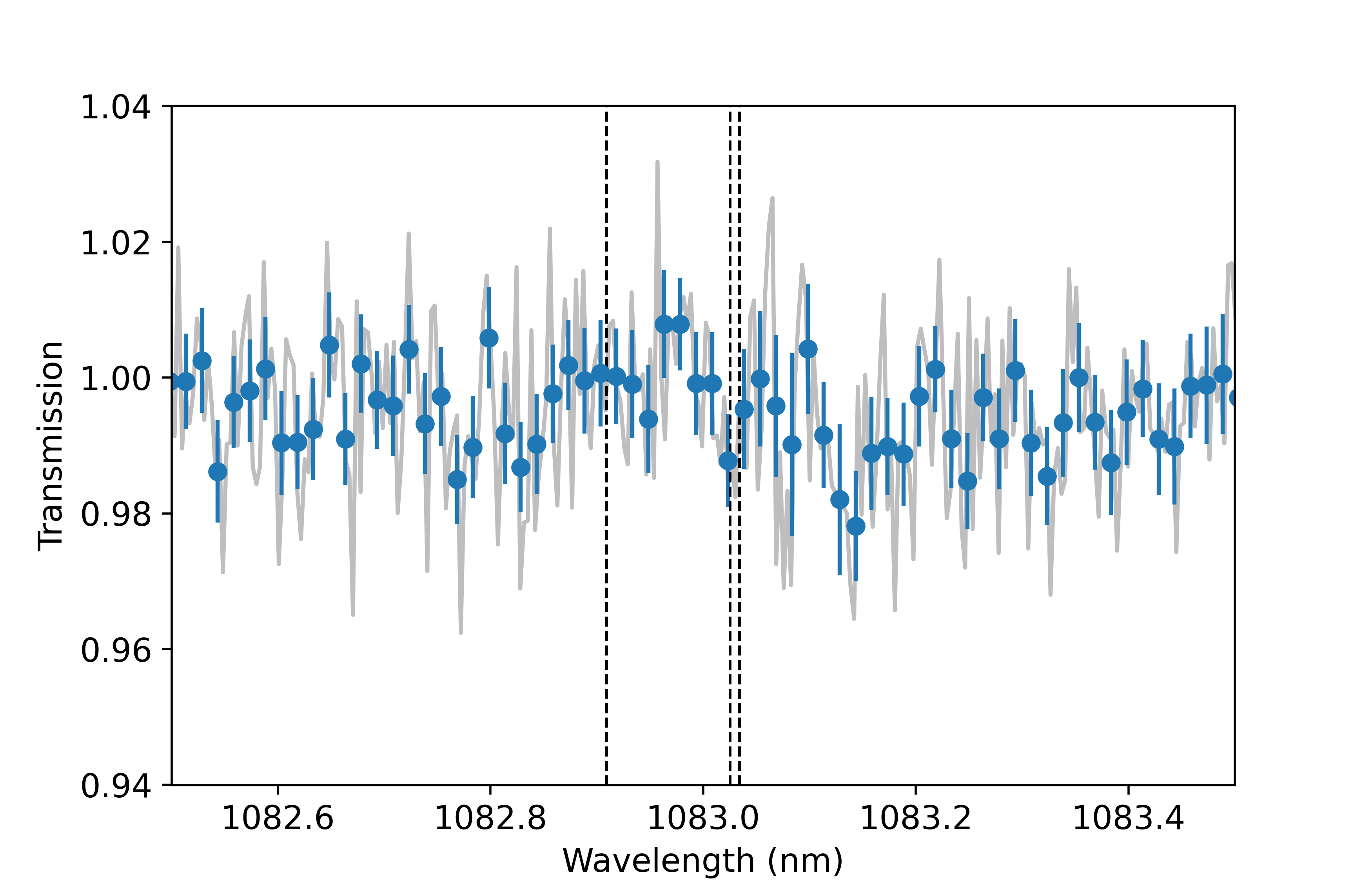}
      \caption{Combined transmission spectrum around the \ion{He}{i} 1083\,nm triplet. The position of the \ion{He}{i} lines are indicated by the vertical dashed lines. The grey line is the original spectrum and the blue dots are the binned spectrum with a bin size of 5 pixels (0.015\,nm).
}
         \label{He-tran}
   \end{figure}
%-----------------------------------------------------------

The transit data observed in 2021 with the Y1029 wavelength setting were used to search for the \ion{He}{i} metastable triplet lines located at air rest wavelengths of 1082.909\,nm, 1083.025\,nm, and 1083.034\,nm. We first removed the telluric and stellar lines using the \texttt{SYSREM} algorithm \citep{Tamuz2005}. We found that with a \texttt{SYSREM} iteration number of two, the telluric and stellar lines around the \ion{He}{i} wavelength are already well removed. Therefore, we chose to use this iteration number for the transmission spectrum calculation. We obtained the transmission spectral matrix by dividing the observed spectra with the computed \texttt{SYSREM} model \citep{Gibson2020}. The spectral matrix was initially calculated in the observer's rest frame. We masked the $\pm$ 5 km\,s$^{-1}$ region around the strong telluric OH emission line at 1083.1322\,nm (air wavelength, \citealp{Oliva2015}) and subsequently shifted the matrix into the stellar rest frame by considering the barycentric Earth's radial velocity ($v_\mathrm{bary}$) and stellar systemic velocity ($v_\mathrm{sys}$ = 35.08 km\,s$^{-1}$, \citealp{Bonfils2018}). 

The final transmission spectral matrix is presented in Fig.~\ref{He-matrix}. There is no obvious absorption signal along the expected trajectory of the \ion{He}{i} lines. It can be seen that the strong telluric OH emission line has little effect on the \ion{He}{i} lines as they only intersect at the end of the transit.  We further computed the combined one-dimensional transmission spectrum. We first shifted the spectra into the planetary rest frame by correcting the radial velocity (RV) caused by planetary orbital motion. The orbital velocity was calculated using a semi-amplitude ($K_\mathrm{p}$) of 102 km\,s$^{-1}$, which was inferred from the planetary orbital parameters \citep{Bonfils2018}. We then averaged all the in-transit spectra to obtain the one-dimensional transmission spectrum (Fig.~\ref{He-tran}). There is also no prominent \ion{He}{i} absorption in the combined transmission spectrum. We provide a 3$\sigma$ He absorption upper limit of 3.6$\%$, computed as three times the standard deviation of the transmission spectrum around the line of interest. Following the method described in the MOPYS project \citep{MOPYS}, we put an upper limit on the observational mass-loss rate of 0.14 M$_{\oplus}$/Gyr.

%When the spectrum is binned every 5 pixels with a wavelength bin of 0.015 nm (the blue points in Fig.~\ref{He-tran}), we obtained a standard deviation of 0.08 for the spectrum, which is regarded as the detection upper limit of the He lines.

\subsection{Search for molecules}

%                                                Two column figure
%----------------------------------------------------------- S_vib
   \begin{figure*}
   \centering
   \includegraphics[width=0.75\textwidth]{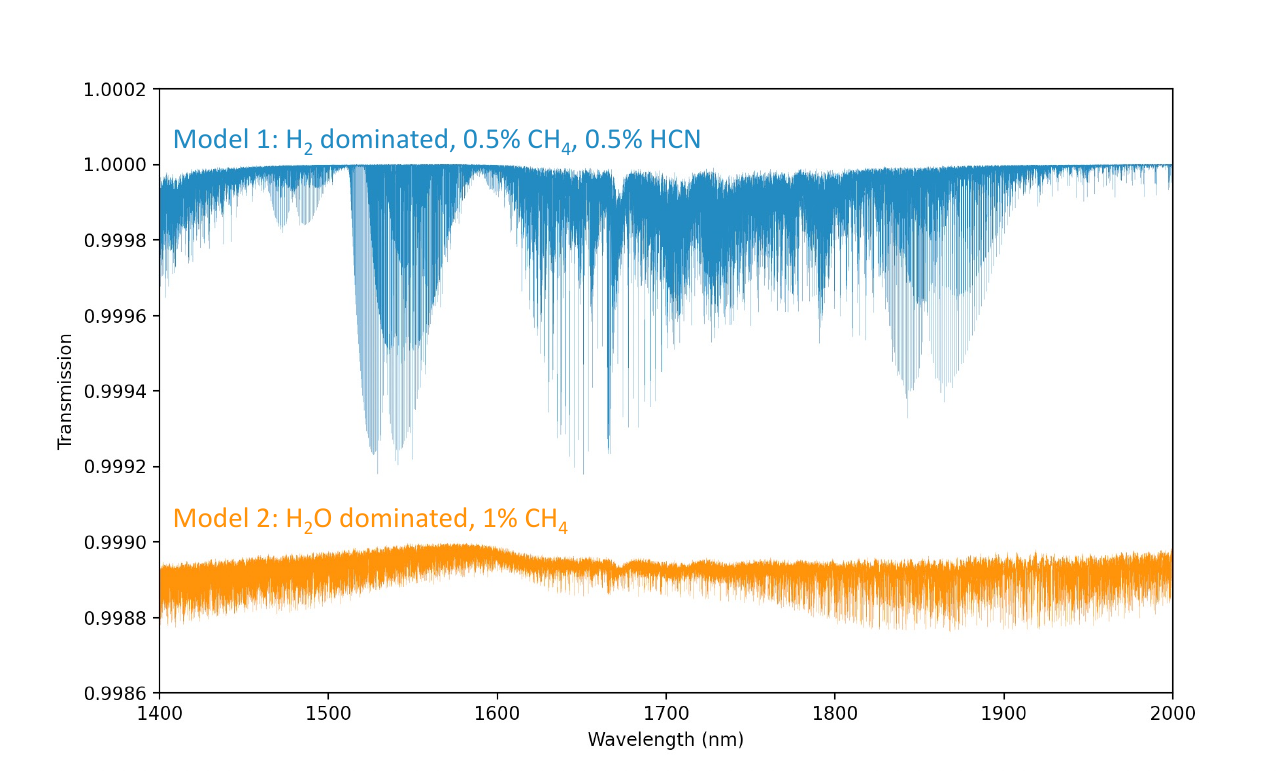}
      \caption{Transmission spectrum models of GJ 1132 b. The spectrum of Model 2 is shifted along the y-axis for clarity.}
         \label{spec-model}
   \end{figure*}
%-----------------------------------------------------------

    \begin{table*}
        \small
        \renewcommand\arraystretch{1.35}
        \caption{Parameters used to calculated the spectral model.}             
        \label{tab-models}      
        \centering          
        \begin{tabular}{l c c c c c c}    
             \hline\hline
             & Composition & Mean molecular weight & Cloud deck pressure & Temperature & Reference \\
             \hline
             Model 1 & 99$\%$ H$_2$, 0.5$\%$ HCN, 0.5$\%$ CH$_4$ & 2.2 & $10^{-3}$\,bar & 480\,K & \cite{Swain2021}\\
             Model 2 & 99$\%$ H$_2$O, 1$\%$ CH$_4$ & 18 & $10^{-1}$\,bar & 530\,K & \cite{May2023}\\
             \hline
        \end{tabular}
        %\tablefoot{\tablefootmark{(a)}{The S/N per pixel was measured at $\sim 6731\,\AA$ for the VIS data and $\sim 12678\,\AA$ for the NIR data.}}
    \end{table*}

%                                                Two column figure
%----------------------------------------------------------- S_vib
   \begin{figure*}
   \centering
   \includegraphics[width=0.8\textwidth]{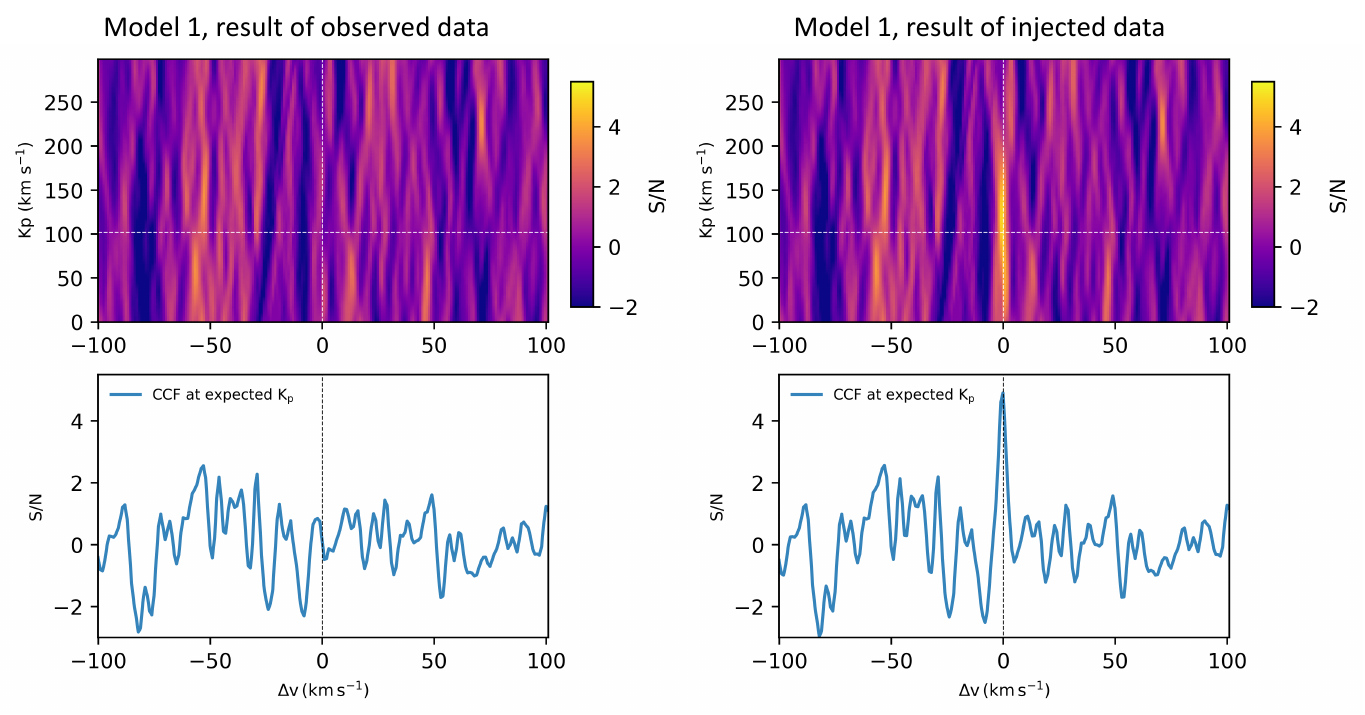}
      \caption{Detection significance map from the two transits observations (left panel). The right panel is the expected detection significance map from the injection test for Model 1, which assumes an H$_2$ dominated atmosphere with CH$_4$ and HCN. }
         \label{kp1}
   \end{figure*}
%-----------------------------------------------------------

%                                                Two column figure
%----------------------------------------------------------- S_vib
   \begin{figure*}
   \centering
   \includegraphics[width=0.8\textwidth]{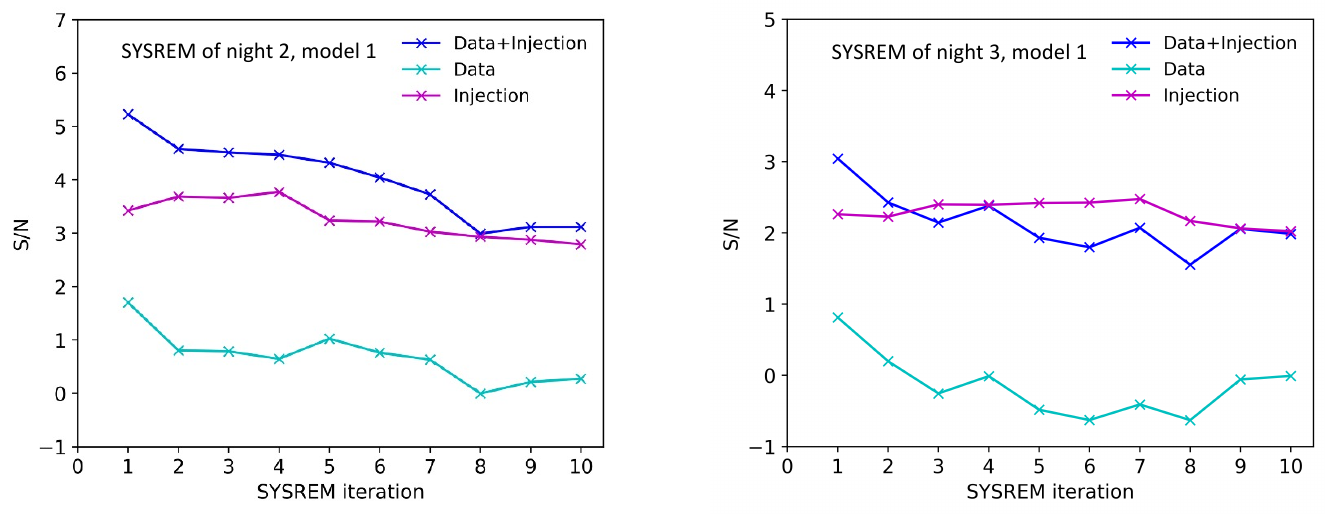}
      \caption{\texttt{SYSREM} evolution with iteration number for the observed data (cyan points), the observed data with injected model (blue points), and the injected model alone (purple points). The results here are for Model 1 (i.e., HCN and CH$_4$ lines).}
         \label{sysrem-model1}
   \end{figure*}
%-----------------------------------------------------------

We used the two transits data taken with the H1559 setting to search for molecular absorption in GJ 1132 b's atmosphere.
To detect the molecular lines in the planetary spectrum, we applied the cross-correlation technique \citep{Snellen2010}. The detailed procedures are similar to the procedures described in \cite{Yan2023}.
We first employed the \texttt{SYSREM} algorithm to remove the stellar and telluric lines imprinted on the observed spectra. Then we calculated the transmission spectral template of GJ 1132 b for two different atmospheric models. The first model has atmospheric parameters similar to the ones reported by \cite{Swain2021}, but we did not include the aerosol-induced Rayleigh scattering slope in the model. Inclusion of the scattering slope would slightly reduce the depth of the spectral lines by about 100 ppm to 200 ppm in the H band. The assumed atmosphere is H$_2$-dominated with 0.5$\%$ HCN and 0.5$\%$ CH$_4$ (volume mixing ratio), along with a cloud deck pressure ($P_\mathrm{cloud}$) of 1\,mbar and an isothermal temperature ($T_\mathrm{iso}$) of 480\,K.
The second model has parameters similar to the retrieved results from the first transit observation in \cite{May2023}. The model is H$_2$O-dominated with 1$\%$ CH$_4$ and we set $P_\mathrm{cloud}$ = 100\,mbar and $T_\mathrm{iso}$ = 530\,K.
The calculation of the spectra was performed with the \texttt{petitRADTRANS} \citep{Molliere2019} tool. We used the line list sources from ExoMol for HCN \citep{Barber2014} and H$_2$O \citep{Polyansky2018} and from HITEMP for CH$_4$ \citep{Hargreaves2020}. The computed spectral model is presented in Figure~\ref{spec-model}. The spectrum of Model 1 consists of both HCN and CH$_4$ lines while the spectrum of Model 2 is mostly H$_2$O lines. The overall line depth of Model 1 is significantly stronger than the line depth of Model 2 because of the lower mean molecular weight of Model 1 (see Table~\ref{tab-models}).

We generated a grid of template spectrum for each model with a step of 1\,km\,s$^{-1}$. The template grid was then cross-correlated with the \texttt{SYSREM}-processed observed data.
We chose to calculate the weighted cross-correlation function (CCF), which uses the inverse of squared noise as the weight of each data point \citep{Gibson2020}. 
These CCFs were then used to compute the so-called $K_\mathrm{p}$-map, which is obtained by adding up all the in-transit CCFs in the planetary rest frame for a given $K_\mathrm{p}$ value. 
The CCFs were initially calculated in the observer's rest frame and we masked the $\pm$ 5 km\,s$^{-1}$ region to avoid contamination by telluric H$_2$O and CH$_4$ lines. When generating the $K_\mathrm{p}$-map, we shifted the CCFs to the planetary rest frame, taking into account $v_\mathrm{bary}$, $v_\mathrm{sys}$ and the planetary orbital motion. To evaluate the detection significance, we calculated the CCF noise value of each $K_\mathrm{p}$-map by computing the standard deviation of the regions with $K_\mathrm{p}$ between 50 to 200 km\,s$^{-1}$ and RV between -150 to -50 km\,s$^{-1}$ or 50 to 150 km\,s$^{-1}$. The detection significance map was subsequently obtained by dividing the $K_\mathrm{p}$-map by the corresponding noise value.
We combined the $K_\mathrm{p}$-maps from the two nights' observation to achieve a higher S/N.

\begin{figure*}
   \centering
   \includegraphics[width=0.8\textwidth]{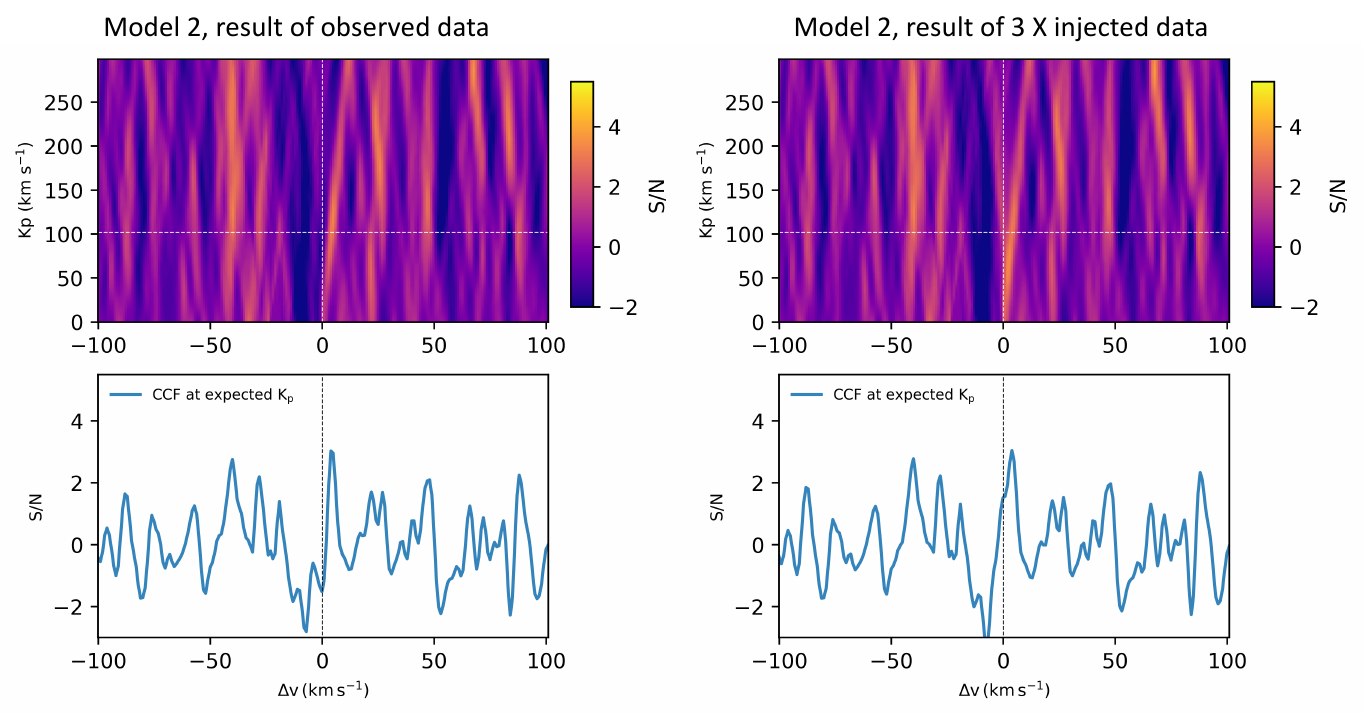}
      \caption{Same as Fig.~\ref{kp1}, but for the case of Model 2 (H$_2$O dominated atmosphere with CH$_4$). Here the injected model spectrum is enhanced by 3 times.}
         \label{kp2}
   \end{figure*}

\begin{figure*}
   \centering
   \includegraphics[width=0.8\textwidth]{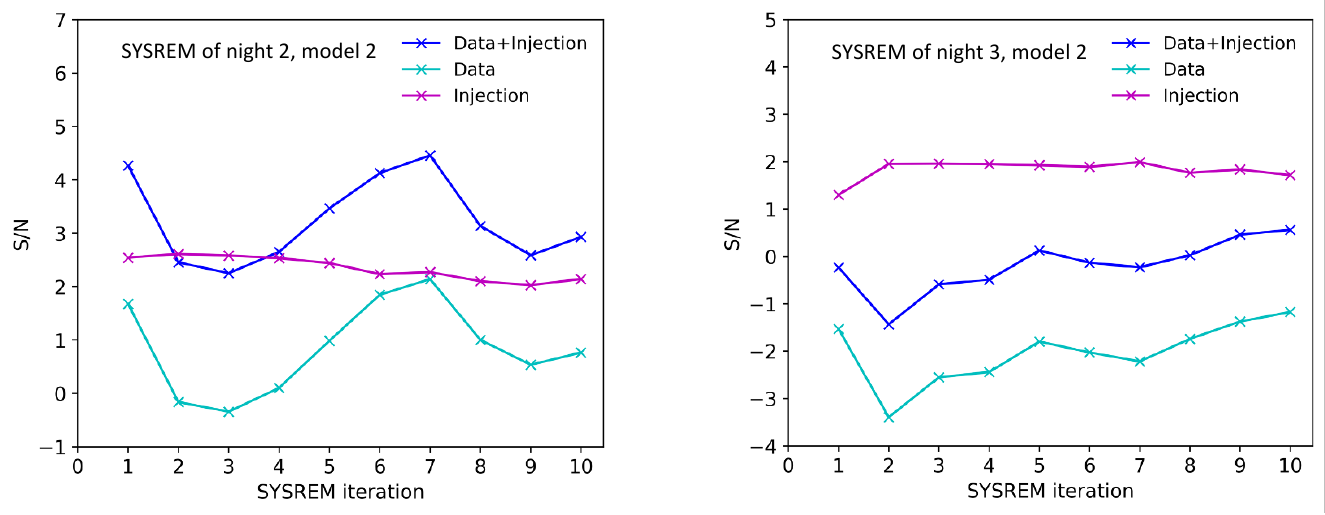}
      \caption{Same as Fig.~\ref{sysrem-model1} but for Model 2 (i.e., mainly H$_2$O lines). The injected model is enhanced by 3 times comparing to the original model in Fig.~\ref{spec-model}.}
         \label{sysrem-model2}
   \end{figure*}

We were not able to detect the HCN and $\mathrm{CH_4}$ lines in Model 1 at the expected $K_\mathrm{p}$ of GJ 1132 b (102 km\,s$^{-1}$). Fig.~\ref{kp1} presents the combined $K_\mathrm{p}$-map of the two nights' observation.
We have tested different \texttt{SYSREM} iteration numbers from 1 to 10, but there is no significant signal on the $K_\mathrm{p}$-map (Fig.~\ref{sysrem-model1}). For Model 2, there is also no prominent signal at the expected $K_\mathrm{p}$ (Fig.~\ref{kp2} and Fig.~\ref{sysrem-model2}), indicating that we did not detect the H$_2$O signal in Model 2. Searches with simple models with individual molecules were equally unsuccessful.

To investigate whether our data are capable of detecting the atmospheric signal as presented in Model 1 and Model 2, we injected the model spectra into the observed data. The injection was only performed for the in-transit spectra and we shifted the model spectra with RVs corresponding to the planetary orbital motion. We then reduced the model-injected data in the same way as described above. For Model 1, the final $K_\mathrm{p}$-map and the detection significance at different \texttt{SYSREM} iteration numbers are presented in Figs.~\ref{kp1} and \ref{sysrem-model1}. When the two nights data are combined, a clear signal reaching 5 $\mathrm{\sigma}$ appears at the expected $K_\mathrm{p}$.
Therefore, our data are, in principle, capable of detecting the HCN+$\mathrm{CH_4}$ signal in Model 1 if the species are indeed present in the atmosphere of GJ 1132 b. 
%\textbf{However, since we did not include aerosols in our model and did not account for the uncertainties in chemical abundances reported in \cite{Swain2021}, we cannot completely rule out the atmospheric scenarios presented in \cite{Swain2021}.}
We also tested the scenario with aerosol scattering as reported in \cite{Swain2021}, and the strength of the injected signal becomes weaker but still reaches 3.5 sigma. Therefore, we can largely rule out the atmospheric scenarios presented in \cite{Swain2021}.
For Model 2, we were not able to detect the injected data even when we enhance the line depth by three times (Fig.~\ref{kp2} and \ref{sysrem-model2}). Thus, the S/N of our data is not high enough to detect the weak H$_2$O signal as presented in Model 2.

We also tested the method described in \cite{Cheverall2023} to optimize the best \texttt{SYSREM} iteration number. We computed the injection-only CCF by subtracting the model-injected CCF with the observation-only CCF. The final S/N at different \texttt{SYSREM} iterations of the injection-only CCF is presented as purple lines in the \texttt{SYSREM} evolution figures. The iteration numbers that provide the highest S/N values of the injection-only CCF were chosen to compute the $K_\mathrm{p}$-maps in Figs.~\ref{kp1} and \ref{kp2}. The injection-only CCF also indicates whether the injected signal can be recovered. For example, the injected Model 1 spectrum can be recovered at the 4$\mathrm{\sigma}$ level for the night 2 data. On the other hand, the three-times enhanced Model 2 spectrum yields a signal below 3$\mathrm{\sigma}$ for the night 2 data.

%                                                Two column figure
%----------------------------------------------------------- S_vib
%-----------------------------------------------------------

%                                                Two column figure
%----------------------------------------------------------- S_vib
%-----------------------------------------------------------

\subsection{Atmospheric retrieval}
To constrain the upper limits of the chemical species, we further performed an atmospheric retrieval with the CRIRES+ data. 
Since our data is not capable of detecting chemical species for a H$_2$O-dominated atmosphere, we only performed the retrieval for the case of an H$_2$-dominated atmosphere. 

The retrieval method is similar to the method in \cite{Yan2020, Yan2023}. The input data is the observed spectral matrix after the removal of the stellar and telluric lines by \texttt{SYSREM}. This data matrix is in the observer's rest frame.
Here a \texttt{SYSREM} iteration number of 4 and 7 were used for night 2 and night 3, respectively. The model transmission spectrum is generated with \texttt{petitRADTRANS}, assuming an H$_2$-dominated atmosphere with a cloud deck at $10^{-3}$\,bar and an iso-thermal temperature of 530\,K. We included CH$_4$, HCN and H$_2$O in the retrieval and set uniform priors for their volume mixing ratios from -10 to -1 (in logarithm). The mean molecular weight of the atmosphere is fixed to 2.2.
Then we generated a model spectral matrix with the same dimensions as the observed data. For each in-transit model spectrum, we shifted it with a RV corresponding to the expected position of the planetary lines in the observer's rest frame, meaning that the planetary orbital RV, $v_\mathrm{bary}$, and $v_\mathrm{sys}$ were all considered. Here we fixed the $K_\mathrm{p}$ to 102\,km\,s$^{-1}$. The parameter setup is summarized in Table \ref{tab-retrieval}.
The model matrix is further processed with a fast \texttt{SYSREM} filter as suggested by \cite{Gibson2022}. Then both the observed data matrix and the model matrix were filtered with a Gaussian high-pass filter with a $\sigma$ of 31 points. 
We subsequently compared the two matrices with the \texttt{emcee} tool \citep{Mackey2013} to evaluate the model parameters in a Markov chain Monte Carlo (MCMC) approach. We run the MCMC for 20000 steps with 200 walkers and set the first 10000 steps as burn-in. 
 
\begin{figure*}
   \centering
   \includegraphics[width=0.8\textwidth]{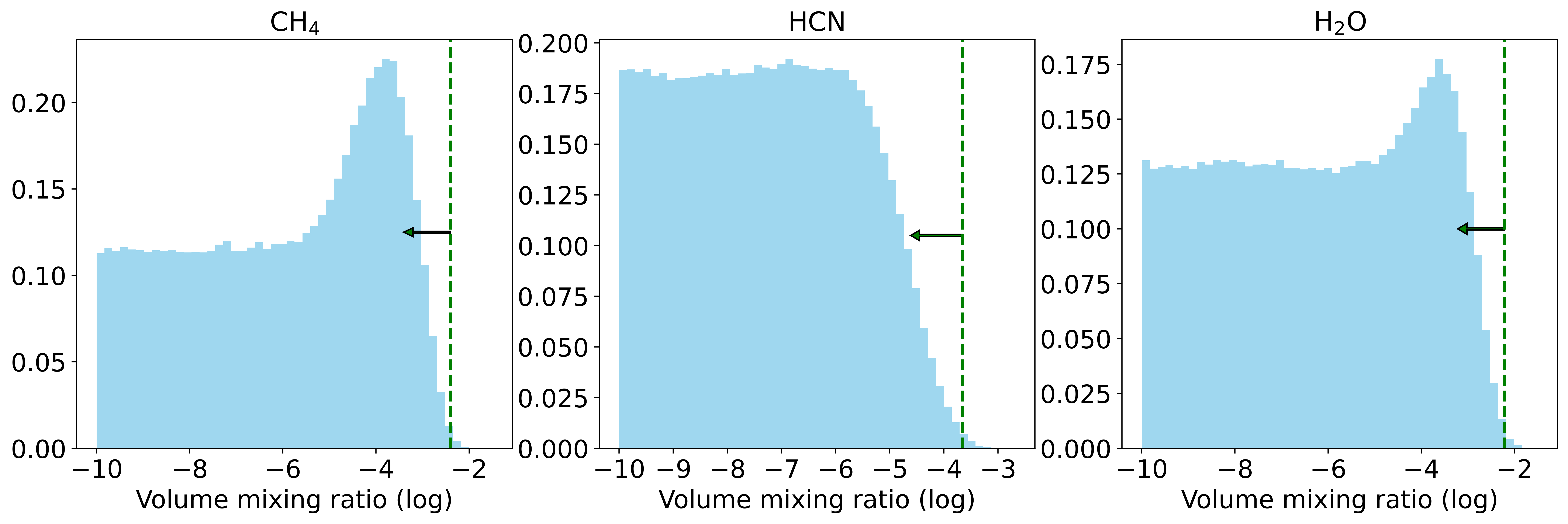}
      \caption{Posterior distributions of the chemical mixing ratios from the retrieval of the observed transmission spectrum. The atmosphere model is assumed to be H$_2$-dominated in the retrieval. The green dashed lines indicate the 3$\sigma$ upper limits. }
         \label{retrieval}
   \end{figure*}

The retrieved result indicates non-detection of the chemical species, which is consistent with the cross-correlation result. Both the retrieval and the cross-correlation results hint that the atmosphere of GJ 1132 b is likely not H$_2$ dominated with a substantial amount of CH$_4$, HCN, or H$_2$O.
According to the posterior distributions (Fig.~\ref{retrieval}), we derived the 3$\sigma$ upper limits for the volume mixing ratios of CH$_4$, HCN and H$_2$O as 10$^{-2.4}$, 10$^{-3.6}$, and 10$^{-2.2}$ and the one-$\sigma$ upper limits as 10$^{-3.7}$, 10$^{-5.5}$ and 10$^{-3.7}$.
These retrieved upper limits show the capability of CRIRES+ in detecting chemical species in rocky exoplanets, if the atmosphere is H$_2$ dominated.

%
%                                             Simple A&A Table
%-----------------------------------------------------------
\begin{table}
\small %%smaller font size
\caption{Retrieval setup.}             % title of Table
\label{tab-retrieval}      % is used to refer this table in the text
\centering                          % used for centering table
\begin{threeparttable}
        \begin{tabular}{l c c}        % centered columns (4 columns)
        \hline\hline \noalign{\smallskip}                 % inserts double horizontal lines
                Parameter & Retrieved value & Boundaries \\     % table heading
        \hline     \noalign{\smallskip}                   % inserts single horizontal line
  \rule{0pt}{2.5ex} log (CH$_4$) &  $-5.7_{-2.9}^{+2.0}$ & -10 to -1  \\ 
    \rule{0pt}{2.5ex} log (HCN)  & $-7.3\pm1.8$ & -10 to -1 \\  
  \rule{0pt}{2.5ex} log (H$_2$O) & $-6.1_{-2.6}^{+2.4}$ & -10 to -1 \\   
  \rule{0pt}{2.5ex} Cloud deck & fixed to $10^{-3}$ bar &  \\ 
  \rule{0pt}{2.5ex} Mean molecular weight &  fixed to 2.2 \\  
  \rule{0pt}{2.5ex} Temperature & fixed to 530\,K &  \\  
  \rule{0pt}{2.5ex} $K_\mathrm{p}$ & fixed to 102 km\,s$^{-1}$ & \\ 
\noalign{\smallskip}  \hline                                   %inserts single line
        \end{tabular}
\end{threeparttable}      
\end{table}
%-----------------------------------------------------------

%                                                Two column figure
%----------------------------------------------------------- S_vib
%-----------------------------------------------------------

\section{Discussion and Conclusions}

The immediate aim of our work was to determine the presence or absence of HCN, CH$_4$, and/or H$_2$O in GJ 1132 b's atmosphere, as well as the probable origin and evolution of such atmosphere. The HST HCN detection by \citet{Swain2021} was highly constraining and results in geochemical outgassing models invoking mantle material with C/O $\approx3$ and H$_2$O mixing ratios $< 1$~ppm. Previous observations with ground-based photometers and low-resolution spectrographs favor alternative scenarios with no thick atmosphere or with a higher H$_2$O mixing ratio. These predictions could be decisively tested with a C/O inventory based on CRIRES+ spectroscopy in the $\sim$1--5 microns interval that include strong bands of H$_2$O, CH$_4$, and CO that are less obscured by the aerosol opacity dominating shorter wavelengths in the transmission spectrum of GJ 1132 b.  Unfortunately we are unable to determine any molecular mixing ratios from our CRIRES+ observations, given our lack of atmospheric signatures detections. 

%\textbf{[Comments from Fei: Our CRIRES+ results can confidentially rule-out the scenario from \citep{Swain2021}, i.e. an H2-dominated atmosphere without thick cloud. So, using the non-detection of H2O as an evidence to support the \citep{Swain2021} scenario may be a bit contradictory. Also, our the 1-sigma upper limits of H2O is 10$^{-3.7}$, which is still away from the required a few ppm.]} Searching for H$_2$O in GJ 1132 b provides a strong test to the hypothesis that ultra-reduced magma is responsible for the outgassed atmosphere. The coupled geochemical/atmospheric modeling results used to interpret the Hubble transmission spectra \citep{Swain2021} predict that GJ 1132 b is a dry, desiccated world; the presence of H$_2$O in quantities substantially greater than a few ppm would challenge the hypothesis that the outgassing magma has an ultra-reduced composition. Testing the ultra-reduced magma hypothesis is important not only for our understanding of the origin of the atmosphere on GJ 1132 b, but also because it has broad implications in the context of how GJ 1132 b formed, and it raises the question of whether mantle ingassing of $H$ routinely produces desiccated terrestrial worlds. The non-detection of H$_2$O by CRIRES+ \textcolor{red}{and the upper limits established by our injection tests seem to support this hypothesis. CHECK}

The detection of H$_2$O -- or placing strict upper limits to its volume mixing ratio -- in GJ 1132 b can potentially provide a strong test to the hypothesis that ultra-reduced magma is responsible for the outgassed atmosphere. The coupled geochemical/atmospheric modeling results used to interpret the HST transmission spectra \citep{Swain2021} predict that GJ 1132 b is a dry, desiccated world; the presence of H$_2$O in quantities substantially greater than a few ppm would challenge the hypothesis that the outgassing magma has an ultra-reduced composition. Testing the ultra-reduced magma hypothesis is important not only for our understanding of the origin of the atmosphere on GJ 1132 b, but also because it has broad implications in the context of how GJ 1132 b formed, and it raises the question of whether mantle ingassing of $H$ routinely produces desiccated terrestrial worlds. Unfortunately, our 3$\sigma$ upper limits for H$_2$O of 10$^{-2.2}$ is much larger than the required, a few ppm, upper limit to consider such an atmosphere as desiccated. On the other hand, our CRIRES+ results can largely rule-out the scenario from \citet{Swain2021}, i.e. an H$_2$-dominated atmosphere without thick cloud.

%\textbf{[Comments from Fei: we have one transit in Y1029 for He, so this paragraph needs re-wording].} We note that CRIRES$^+$ time was also awarded to explore the presence of He I in the planetary atmosphere, in the Y1029.31 wavelength setting but could not be scheduled. 

The metastable triplet of He I has been established as one of the best tracers for planetary mass loss accessible to ground-based telescopes \citep{Nortmann2018}. One of the most important aspects in interpreting \cite{Swain2021} results is whether the current epoch atmosphere is a remnant of the primordial H/He envelope or a regenerated atmosphere. 
The detection of a He I extended atmosphere would have suggested that GJ 1132 b has undergone orbital migration relatively recently (past ~100 Myr) and thus retains a portion of the primordial H/He envelope. Our non-detection of He I in the planetary atmosphere supports the hypothesis that the present-epoch atmosphere is secondary, but it is also consistent with an airless/tenuous atmosphere. However, we must note that the detection of He I is highly dependent on the extreme ultraviolet (EUV) flux received from the host star, so it is possible that our non-detection responds to a low irradiance values at these wavelengths \citep{Lampon2023}.

%Thus, ground-based observation of He I absorbtion have the potential to determine if GJ 1132b has an extended primordial atmosphere or not. 

The results from the first JWST visit to GJ 1132 b \citep{May2023} were consistent with a H$_2$O-dominated atmosphere with CH$_4$ and N$_2$O, but also with stellar contamination from unocculted starspots. The second visit, however, resulted in a flat spectrum, with the two observations separated by only 8 days, without a satifactory explanation. It is important to note that high-resolution data can in principle distinguish between the atmospheric signal and the often varying stellar contamination using the RV signature (i.e., the atmospheric signature has a $K_{\rm p}$ of 102 km/s, while the stellar contamination does not). Unfortunately, according to our recovery-injection test, two transits of CRIRES+ are not sufficient to detect the H$_2$O spectral feature if the atmosphere has a high mean molecular weight such as in the H$_2$O-dominated scenario. However, these retrieved upper limits show the capability of CRIRES+ in detecting chemical species in rocky exoplanets, if the atmosphere is H$_2$ dominated. In the future, the capabilities of instruments such as ANDES at the Extremely Large Telescope will be able to probe the atmospheres of small rocky worlds \citep{Palle2023} with high mean molecular weight, and to distinguish between the variability introduced by stellar activity and/or the planetary atmosphere.

%\textbf{Comments from Fei: Maybe add some discussion on the H$_2$O-dominated scenario from the visit-1 of JWST?
%Point 1: our high-resolution data can in principle distinguish between the atmospheric signal and the stellar contamination using the RV signature (i.e., the atmospheric signature has a Kp of 102 km/s, while the stellar contamination does not.)
%Point 2: however, according to our recovery-injection test, two transits of CRIRES+ is not sufficient to detect the H$_2$O spectral feature if the atmosphere has a high mean molecular weight such as the H$_2$O-dominated scenario.} 

%Accumulating more transits observation with CRIRES+/VLT or using the future ANDES/ELT will be able to distinguish between the atmospheric signal and the stellar contamination.

%\textcolor{red}{Should we discuss the possibility of planetary variability to explain the lack of detections? JWST had two unconsistent transits separated by 8 days. Ours are separated by 18 days, and consistent.} \textbf{Comments from Fei: this may be difficult, since none of our CRIRES+ transits provides a signal higher than 3-sigma, and also the JWST transit 1 result is beyond the detection limit of our CRIRES+ data. So it may not be convincing to attribute our lack of detections to planetary variability.}

%\textcolor{red}{One or two closing sentences}

\begin{acknowledgements}

We acknowledge financial support from the Agencia Estatal de Investigaci\'on of the Ministerio de Ciencia e Innovaci\'on MCIN/AEI/10.13039/501100011033 and the ERDF “A way of making Europe” through project PID2021-125627OB-C32, and from the Centre of Excellence “Severo Ochoa” award to the Instituto de Astrofisica de Canarias. 
F.Y. acknowledges the support by the National Natural Science Foundation of China (grant No. 42375118). 
G.M. acknowledges financial support from the Severo Ochoa grant CEX2021-001131-S and from the Ram\'on y Cajal grant RYC2022-037854-I funded by MCIN/AEI/ 10.13039/501100011033 and FSE+.
P. A. Miles-P\'aez acknowledges financial support from  the grant RYC2021-031173-I funded by MCIN/AEI/ 10.13039/501100011033 and by the 'European Union NextGenerationEU/PRTR'.
TM acknowledges financial support from the Spanish Ministry of Science and Innovation (MICINN) through the Spanish State Research Agency, under the Severo Ochoa Program 2020-2023 (CEX2019-000920-S).
We thank Guo Chen for useful discussions on model calculations.
M.S. acknowledges the support of the PRIN INAF 2019 through the project “HOT-ATMOS” and INAF GO Large Grant 2023 GAPS-2.

\end{acknowledgements}


% for the bibliography, at the end
\bibliographystyle{aa} % style aa.bst

\bibliography{CRIRES-refer}

\end{document}